\documentstyle[prd,aps,epsfig,preprint]{revtex}
\tighten

\begin{document}
\thispagestyle{empty}
\baselineskip24pt
\draft
\begin{center}
{\large\bf Alternative Form of the Parity-Violating 
Current for the Hyperon Weak Radiative Decays and Hara Theorem}
\end{center}
 \begin{center}
{Elena~N.~Bukina$^a$, Vladimir~M.~Dubovik$^a$ and
Valery~S.~Zamiralov$^{b}$}
\hbox{\it $^a$Joint Institute for Nuclear Research,
          141980 Dubna, Moscow region, Russia} 
\hbox{\it $^b$ D.V. Skobeltsyn Institute of Nuclear Physics,
Moscow State University, Moscow, Russia}
\end{center}
\date{\today}
\begin{abstract}
It is shown that upon considering an alternative form 
of a parity-violating part of the transition electromagnetic current
it is possible to reformulate Hara theorem in a way that it does not forbid 
any more nonzero asymmetry in the hyperon weak radiative decays
$ \Sigma^{+}\rightarrow p+\gamma $ and $ \Xi^{-}\rightarrow 
\Sigma^{-}+\gamma $ in the limit of exact  $SU(3)_{f}$ symmetry
thus resolving a contradiction with the data and maybe revealing
hitherto unseen transition toroid dipole moments. 
A result is consistent with the traditional one on the single-quark weak
radiative transition models. 
We have also reproduced Vasanti formula at the quark level. 
As for two-quark weak radiative transitions we have found that the
important part of it contains also a toroid dipole moment
contribution which seems to be an intrinsic reason of the apparent 
contradiction between the Hara theorem conclusion and quark model 
results for hyperon weak radiative decays.
\vskip 3mm
\noindent
{\bf PACS} number(s): 11.30.Hv, 13.30.-a, 13.40.Hg, 14.20.Jn
\end{abstract}


%

\section{Introduction}
\setcounter{equation}{00}
$\quad$ The weak radiative decays have been first analyzed theoretically
about forty years ago \cite{Behr}-\cite{Pati}. It was
envisaged already in \cite{Salam} a possibility to understand it in the 
framework of pole model similar to the nonleptonic hyperon decays. 
Even earlier estimations of the decay rates were made basing on the pion 
photoproduction amplitudes \cite{Nish}. At the same time two 
experiments were performed \cite{Bo,Te}, where the first events of the decay
$\Sigma^{+} \rightarrow p + \gamma $ were found (in all 7 events).   
Unitary symmetry arrived, more elaborated schemes appeared 
(see, e.g., \cite{Pak}) and a theorem was proved by Hara that
decay asymmetry in the charged hyperon weak radiative decays 
$\Sigma^{+} \rightarrow p + \gamma $ 
and $ \Xi^{-} \rightarrow \Sigma^{-} + \gamma $ should  vanish 
in the limit of exact $SU(3)_{f}$
\cite{Hara} while it can be non zero for the neutral decays
$(\Sigma^{0},\Lambda) \Rightarrow n + \gamma $ and
$ \Xi^{0} \rightarrow (\Sigma^{0},\Lambda) + \gamma $.
Since experimental discovery of a large negative asymmetry 
in the radiative decay $ \Sigma^{+} \rightarrow p + \gamma $  \cite{Ger}, 
confirmed  later \cite{PDG} (see the Table 1),
the explanation of the  net contradiction 
between  experimental results and 
the Hara theorem prediction ``has constituted a constant challenge to 
theorists'' \cite{ZenM}. The contradiction looks even more strange in the light of 
existing estimates for asymmetry values of other hyperon radiative decays
which though not measured so precise as that of $\Sigma^{+} \rightarrow p + \gamma $ 
seem to be of the same order of magnitude \cite{PDG}.
It would eventually require large $SU(3)_{f}$ symmetry breaking terms but they
hardly can be large enough due to the well-known
Ademollo-Gatto theorem \cite{AG} in order to be able to account 
for this difficulty.
   
Another puzzle is related to inconsistency between $ SU(3)_{f}$ symmetry
and quark model predictions for the asymmetry value in the hyperon weak 
radiative decays. 
Indeed, the Hara theorem was formulated at the hadron level in terms of the
$ SU(3)_{f}$ baryon wave functions. It would seem natural if in the 
framework of a quark model one arrived 
at a similar result with some minor deviations.
But quark models while more or less succeding in describing experimental
data on branching ratios and asymmetry parameters (see, e.g., a review 
\cite{ZenM} and citations therein)
did not reproduce the Hara claim without making vanish
all asymmetry parameters in the $SU(3)_{f}$ symmetry limit.
The origin of this discrepancy is not clear up to now although many
authors have investigated this problem thoroughly 
\cite{Ryaz}-\cite{Dmit} (see also \cite{ZenM} for 
very complete list of publications)
and is a real puzzle as similar calculations say of baryon magnetic moments 
are known to be rather consistent at the quark and hadron level. 
A single quark weak radiative transition $ s \rightarrow d + \gamma $
was calculated in the framework of the standard Weinberg-Salam model 
with diagram technique \cite{Gad}, and it was shown that its
decay asymmetry is proportional to the ratio $ (m_{s}-m_{d})/(m_{s}+m_{d})$
where $ m_{q} \, (q=d,s)$ is the current quark mass. The same result was 
obtained earlier in \cite{Vasa} on invoking chiral symmetry arguments. 
So the parity-violating quark transition amplitude goes to zero in the chiral 
symmetry limit. But these results cannot be, generally speaking, immediately 
translated to the hyperon weak radiative decays, as one should first 
relate current quark evaluations to baryon picture which is far from trivial. 
But the single-quark contributions proved to be too small to account for
the observed decay rates. Even penguin diagram contributions are
not strong enough to enhance an effective $ s \rightarrow d+\gamma$.
Two-quark weak radiative diagrams (those with exchange
of W-boson between two quarks, one of them emitting a photon, 
the third quark being a spectator) were shown to give in general 
nonzero contribution to 
decay asymmetry of all 5 hyperon radiative decays  including 
$\Sigma^{+} \rightarrow p + \gamma $ \cite{Ryaz}-\cite{Verma}. 
(Note that the 6th decay $ \Xi^{-} \rightarrow \Sigma^{-} + \gamma $
cannot proceed via the two-quark W-exchange diagram). Moreover their
contribution into the decay rate proved to be important. But the problem of
the mutual inconsistency between two-quark diagram results and 
Hara theorem prediction still persists.

In what follows we shall try to show that a discrepancy between the Hara 
theorem prediction and experimental result for the asymmetry in the decay
$\Sigma^{+} \rightarrow p + \gamma $  may be overcome due to  alternative
possibility of multipole parametrization of the axial 
transition electromagnetic current which include not only dipole transition 
moment but also contribution of the toroid dipole moment \cite{Chesh,DT}. 
As it has been pointed recently \cite{Bukina}, toroid dipole moment naturally
arrives in the parity-violating (PV) part of the transition weak radiative
matrix element and leads to reformulation of the Hara theorem. We shall
also show that the Vasanti result as to the single-quark weak radiative transition
\cite{Vasa} is reproduced in our scheme while going from hadron to 
quark level. 
Two-quark  weak radiative transition are considered also and a source of 
discrepancy between the results of $ SU(3)$ symmetry approach of \cite{Hara} 
and that of quark models \cite{Ryaz,Sharma} and others seems 
to be established.

\section{ Kinematics of the hyperon radiative decay process }
\setcounter{equation}{00}


The two-body radiative decay amplitude 
$A (B_{1} \rightarrow B_{2}+ \gamma)$ 
in the rest frame of $ B $ , usually is written in the form \cite{Behr}
\begin{equation}
A (B_{1} \rightarrow B_{2} \gamma)=i \sqrt{ \frac{m_{N}}{4 \pi k_{0}E_{N}}}
\overline{u}_{2}(p_{2})(C+D \gamma_{5}) \sigma_{\mu \nu} k^{ \nu} u_{1}(p_{1})
\delta^{4}(p_{1}-p_{2}-k),
\end{equation}
which preserves automatically gauge invariance condition.
Here $ u_{1,2} $ are Dirac spinors of the baryons $ B_{1,2}$ with masses
$  m_{1,2}$, respectively, while $ C$ and $ D$ are parity-conserving (PC)
and parity-violating (PV) amplitudes, correspondingly.
The photon momentum value $ k_{ \nu}$ is entirely determined in the rest frame
of $ B$ by baryon masses $ m_{1,2}$,
The angular distribution of the photon reads
$$ W(\theta)=\frac{1}{16 \pi}\left( \frac{m_{1}^{2}-m_{2}^{2}}{m_{1}}
\right)^{3}
(|C|^{2}+|D|^{2})[1+\alpha(\widehat s \widehat p)], $$
where $ \widehat s$ is the  polarization vector of $ B$ in its rest frame,
$ \theta $ is the angle between the direction of polarization of $ B$ and the 
momentum of $ B'$,
and $ \widehat p$ being the direction of the momentum of $ B'$, while
$ \alpha$ is the decay asymmery parameter.
In terms of $ C$ and $ D$ the decay asymmetry
is written as \cite{Behr}

\begin{equation}
\alpha=\frac{2Re(C^{*}D)}{|C|^2+|D|^2}.
\label{alpha}
\end{equation}

The corresponding decay rate is given by
\begin{equation}
R=\frac{1}{8 \pi}\left(\frac{m_{1}^{2}-m_{2}^{2}}{m_{1}}\right)^{3}
(|C|^2+|D|^2).
\label{rate}
\end{equation}
Experimental data on decay rates in the form of corresponding
branching ratios $ BR=R(B \rightarrow B'+
\gamma)/R(total))$ and asymmetry parameters of the
relevant decays \cite{PDG} are placed in the Table 1. 


\section{ Towards the multipole parametrization of transition vector
current}
\setcounter{equation}{00}

   
 The transition vector current of the two particles with spin $1/2$ and 
 parity can be expressed in terms of 5 Lorentz structures
 $ \gamma_{\mu}$, $P_{\mu}$,  $k_{\mu}$,  $\sigma_{\mu \nu} k^{\nu}$ and 
 $ \sigma_{\mu \nu} P^{\nu}$, where 
 $ P_{\mu}=(p_{1}+p_{2})_{\mu}$, $k_{\mu}=(p_{1}-p_{2})_{\mu}$,
 $\sigma_{\mu \nu}=(i/2)\left[\gamma_{\mu},\, \gamma_{\nu}\right]$.
Upon using current conservation condition one is left with two Lorentz
structures, so the effective transition parity-conserving current reads
in one of the forms \cite{Chesh}:

\begin{eqnarray}
J^{(V)}_{\mu}(k_{\nu})=\frac{e \eta}{(2 \pi)^{3}}\overline{u}_{2}\left[
\frac{1}{M^2}\left(k^{2}_{\lambda}\gamma_{\mu}-\widehat{k}k_{\mu}
\right)F_{1}(k^{2}_{\lambda})
+\frac{1}{M}\sigma_{\mu \nu}k_{\mu}F_{2}(k^{2}_{\lambda})\right]u_{1}
\label{VD1}
\\
 =\frac{e \eta \gamma^{2}}{(2 \pi)^{3}}
\overline{u}_{2}\left[\frac{1}{M^3}\left(k^{2}_{\lambda}P_{\mu}-
(k_{\nu}P_{\nu})k_{\mu}\right)F_{3}(k^{2}_{\lambda})+
\frac{1}{\gamma^{2} M}\sigma_{\mu \nu}k_{\mu}F_{4}(k^{2}_{\lambda})\right]
u_{1}
\label{VD2}
\\
=\frac{e \eta \gamma^{2}}{(2 \pi)^{3}}\overline{u}_{2}
\left[\frac{1}{M^3}\left(k^{2}_{\lambda}P_{\mu}-
(k_{\nu}P_{\nu})k_{\mu}\right)F_{5}(k^{2}_{\lambda})+\frac{1}{\gamma^{2}
M^{2}}
(k^{2}_{\lambda}\gamma_{\mu}-\widehat{k}k_{\mu})F_{6}(k^{2}_{\lambda})
\right]u_{1}
\label{VD3}
\\
=\frac{e \eta \gamma^{2}}{(2 \pi)^{3}}
\overline{u}_{2}\left[\frac{1}{M^3}
\left(k^{2}_{\lambda}P_{\mu}-(k_{\nu}P_{\nu})k_{\mu}\right)
F^{(e)}(k^{2}_{\lambda})+ \frac{i}{M^2}
\epsilon_{\mu \nu \lambda \sigma}P_{\nu}k_{\lambda}
\gamma_{\sigma}\gamma_{5}
F^{(m)}(k^{2}_{\lambda}) \right]u_{1}.
\label{VD}
\end{eqnarray}
Here $ M=m_{1}+m_{2}$, $\Delta m=m_{1}-m_{2}$,
$\eta = \sqrt{1-\Delta m^{2}/M^{2}} $  and $\gamma = 1/\sqrt{1-
k^{2}_{\lambda}/M^2}$ is the Lorentz-factor.
 
The form factors of each current configuration can be expressed in terms
of another one upon using Gordon identities in the form
\cite{Chesh,BDK} (only two of them are independent~):

\begin{eqnarray}
\overline{u}_{2}\left\{ k^{2}_{\lambda}\sigma_{\mu \nu}k_{\nu}+
M(k^{2}_{\lambda}\gamma_{\mu}-\widehat{k}k_{\mu})-\left[k^{2}_{\lambda}P_{\mu}-
(k_{\nu}P_{\nu})k_{\mu}\right] \right\}u_{1}=0,
\nonumber \\
\overline{u}_{2}\left\{ik^{2}\epsilon_{\mu \nu \lambda \sigma}P_{\nu}k_{\lambda}
\gamma_{\sigma}\gamma_{5}+M \left[k^{2}_{\lambda}P_{\mu}-
(k_{\nu}P_{\nu})k_{\mu}\right]+(k^{2}_{\lambda}-M^{2})
(k^{2}_{\lambda}\gamma_{\mu}-\widehat{k}k_{\mu})\right\}u_{1}=0,
\nonumber \\
\overline{u}_{2}\left\{(M^{2}-k^{2}_{\lambda})\sigma_{\mu \nu}k_{\nu}+
iM\epsilon_{\mu \nu \lambda \sigma}P_{\nu}k_{\lambda}
\gamma_{\sigma}\gamma_{5}+\left[k^{2}_{\lambda}P_{\mu}-
(k_{\nu}P_{\nu})k_{\mu}\right] \right\}u_{1}=0,
\nonumber \\
\overline{u}_{2}\left\{i\epsilon_{\mu \nu \lambda \sigma}P_{\nu}k_{\lambda}
\gamma_{\sigma}\gamma_{5}+ M \sigma_{\mu \nu}k_{\nu}+
(k^{2}_{\lambda}\gamma_{\mu}- \widehat{k}k_{\mu})\right\}u_{1}=0.
\nonumber
\end{eqnarray}
Using the identities we may find out relations between all
parametrizations
Eqs.(\ref{VD1})-(\ref{VD}) and observe some kinematic peculiarities.
Moreover, pursuing the multipole analysis of currents in
spirit of \cite{Sachs,Bart} one can see that in general
form factors do not correspond
to the  definite multipole distributions. Only those of Eq.(\ref{VD})
are in fact multipole ones and may be classified on the complete scheme of
multipole expansion of classical electromagnetic current \cite{Chesh,BDK}.
The full formalism of multipole consideration includes not only moments
but, in principle, an infinite sequence of $2n$-power radius for each
moment. Generally, the latter corresponds to expansion of each
form-factors in series on ${\mbox{\boldmath $k$}}^{2}$\cite{Chesh}.
However, considering two-body decay
the form-factor identification with the parameters mentioned may be done
only in a special reference system, named in \cite{Chesh} by the intrinsic
one (see also \cite{BDK}). This reference system is given by the equality
of kinetic energies (e.k.e.) of both the baryons involved and enables us
to
write the nonrelativistic reduction of Eq.(\ref{VD})
\begin{itemize}
\item
for the electric contribution
\begin{eqnarray}
\frac{\eta \gamma^2}{M}F^{(e)}(k_{\lambda}^{2})&\overline{u}_{2}&
\left(k^{2}_{\lambda}P_{\mu}-(k_{\nu}P_{\nu})k_{\mu}\right) u_{1}
A_{\mu} \nonumber \\
& \stackrel{e.k.e.}{\Longrightarrow} &
F^{(e)}(\Delta m^2 - {\mbox{\boldmath $k$}}^{2}) \varphi^{+}_{2} \varphi_{1}
\left( -{\mbox{\boldmath $k$}}^{2} \Phi - \Delta m \mbox{\boldmath $k$}
\mbox{\boldmath $A$} \right)  \nonumber \\
&\Longrightarrow&
F^{(e)}(\Delta m^2 - {\mbox{\boldmath $k$}}^{2}) \varphi^{+}_{2} \varphi_{1}
\left( {\mbox{\boldmath $\nabla$}}^{2} \Phi +  \mbox{\boldmath $\nabla$}
\dot{\mbox{\boldmath $A$}} \right) \\
&\stackrel{|{\mbox{\boldmath $k$}}| \to 0}{\Longrightarrow}& -
F^{(e)}(\Delta m^2) \varphi^{+}_{2} \varphi_{1}
\mbox{div} {\mbox{\boldmath $E$}} =: - \frac{1}{6}\overline{r^{2}_{Q}} 
\varphi^{+}_{2} \varphi_{1}\rho^{ext},
\nonumber
\end{eqnarray}
\item
for the magnetic contribution
\begin{eqnarray} 
\frac{i\eta \gamma^2}{M}F^{(m)}(k_{\lambda}^{2}) &\overline{u}_{2}&
\epsilon_{\mu \nu \lambda \rho } P_{\nu} k_{\lambda}\gamma_{\rho}
\gamma_{5}
u_{1} A_{\mu}
\nonumber \\
&\stackrel{e.k.e.}{\Longrightarrow} &
\frac{i \gamma}{M}F^{(m)}(\Delta m^2 - {\mbox{\boldmath $k$}}^{2})
\varphi^{+}_{2} \left( \epsilon_{iojk}P_{0}k_{j}\sigma_{k} +
\epsilon_{ij0k}k_{0}\sigma_{k} \right) \varphi_{1} A_{i}
\nonumber \\
&=&  iF^{m}(\Delta m^2 - {\mbox{\boldmath $k$}}^{2}) \varphi^{+}_{2}
\left( -\frac{{\mbox{\boldmath $k$}}^{2} [{\mbox{\boldmath $k$}} \times
{\mbox{\boldmath $\sigma$}}]}{k^{2}_{\lambda}} +
\frac{\Delta m^2 [\mbox{\boldmath $k$} \times
\mbox{\boldmath $\sigma$}]}{k^{2}_{\lambda}} \right) \varphi_{1}
\mbox{\boldmath $A$} 
\nonumber \\
&\stackrel{| \mbox{\boldmath $k$}| \to 0}{\Longrightarrow}&
F^{(m)}(\Delta m^2) \varphi_{2}^{+} \mbox{\boldmath $\sigma$} \varphi_{1}
\mbox{rot} \mbox{\boldmath $A$} =: \mbox{\boldmath $\mu$} 
\mbox{\boldmath $B$}.
\end{eqnarray}
Here $\Phi$ and $\mbox{\boldmath $A$}$ are the external potentials,
$\mbox{\boldmath $E$}$ and $\mbox{\boldmath $B$}$ are the
electric and magnetic fields, respectively, and $\varphi_{1,2}$ are Pauli
spinors of baryons in consideration.
\end{itemize}
Thus we may connect $F^{(e)}(\Delta m^2)$ with the standard multipole
parameter $\overline{r^{2}_{Q}}$, the mean-square radius of charge
density distribution, and consider $F^{(m)}(\Delta m^2)$ as the
projection of magnetic dipole moment on $\mbox{\boldmath $\sigma$}$.
Taking into account the relation between our parametrizations easily to
find
the usual decomposition for the diagonal case $m_{1} = m_{2} = m_{0}$,
e.g.
\begin{eqnarray*}
\overline{r_{q}^{2}} = \frac{3e}{2m_{0}^{2}} F^{(e)}(\Delta m^2) =
\frac{3e}{2m_{0}^{2}}
\left[ F_{5}(0) + F_{6}(0)\right],
\end{eqnarray*}
where $\overline{r_{q}^{2}}$ of the baryon considered is given in
$e/m_{0}^{2}$ unit that corresponds to the normalization factors
in Eqs.(\ref{VD1})-(\ref{VD}).
Instead other parametrizations do not give such simple answer.
It is not a proof of the validity of this very expansion given
by Eq.(\ref{VD}). So, more natural to use the multipole parametrization
for the vector current.
But the situation is more dramatic in the case of the axial-vector
current  
as it will be seen in the next section.

\section{ Towards the multipole parametrization of transition axial
current}
\setcounter{equation}{00}

$\quad$ Let us consider axial electromagnetic transition current of the
two particles with spin $1/2$ and parity.
Its possible form is not unique as the most general expression can be
written in terms of $5$ Lorentz structures:  
$ \gamma_{\mu} \gamma_{5}$, $P_{\mu}\gamma_{5}$, $k_{\mu}\gamma_{5}$,
$\sigma_{\mu \nu} k^{\nu} \gamma_{5} $ and 
$ i \epsilon_{\mu \nu \rho \lambda} \gamma_{\nu}P_{\rho}k_{\lambda} $, where 
$ P_{\mu}=(p_{1}+p_{2})_{\mu}$, $k_{\mu}=(p_{1}-p_{2})_{\mu}$, 
 $\sigma_{\mu \nu}=(i/2)\left[ \gamma_{\mu},\gamma_{\nu} \right] $.
But due to electromagnetic current conservation and generalized
Gordon identities (see, e.g., \cite{Chesh})
(only two of them are independent)
\begin{eqnarray}
\overline{u}_2\{ \Delta m \sigma_{\mu \nu} k_{\nu}+
(k_{\lambda}^{2}\gamma_{\mu}-\widehat{k}k_{\mu})+
i\varepsilon_{\mu \nu \lambda \sigma}P_{\nu}k_{\lambda}\gamma_{\sigma}
\gamma_{5}\} \gamma_{5} u_1=0,
\nonumber \\
\overline{u}_2\{k_{\lambda}^{2}\sigma_{\mu \nu}k_{\nu}+\Delta m
(k_{\lambda}^{2}\gamma_{\mu}-\widehat{k}k_{\mu})+[k_{\lambda}^{2}P_{\mu}-
k_{\lambda}P_{\lambda}k_{\mu}] \} \gamma_{5}u_1=0, 
\nonumber \\
\overline{u}_2\{ik^2\varepsilon_{\mu \nu \lambda \sigma}
P_{\nu}k_{\lambda}\gamma_{\sigma}
\gamma_{5} +(\Delta m^2 - k^2)
(k_{\lambda}^{2}\gamma_{\mu}-\widehat{k}k_{\mu})+
\Delta m [k_{\lambda}^{2}P_{\mu}-
k_{\lambda}P_{\lambda}k_{\mu}] \} \gamma_{5}u_1=0, 
\nonumber \\
\overline{u}_2\{-i\Delta m \varepsilon_{\mu \nu \lambda \sigma}  
P_{\nu}k_{\lambda}\gamma_{\sigma}
\gamma_{5} + (k^2 - \Delta m^2)
\sigma_{\mu \nu} k_{\nu}+ [k_{\lambda}^{2}P_{\mu}-
k_{\lambda}P_{\lambda}k_{\mu}] \} \gamma_{5}u_1=0, \nonumber 
\end{eqnarray}
where $ \Delta m= m_1-m_2 $ and $u_{1,2}$ are the Dirac spinors of 
baryons with masses $m_{1,2}$, this transition current
can be reduced to one of the following forms \cite{Chesh,DT}
\begin{eqnarray}
J^{(A)}_{\mu}(k_{\nu})=\frac{e \eta \gamma}{(2 \pi)^{3}}
\overline{u}_{2}\left[
\frac{1}{M^{2}}
(k^{2}_{\lambda}\gamma_{\mu}-kk_{\mu})G_{1}(k^{2}_{\lambda})+
\frac{1}{M}\sigma_{\mu \nu}k_{\mu}G_{2}(k^{2}_{\lambda})
\right]\gamma_{5}u_{1},
\label{JA}     \\[24pt]
J^{(A)}_{\mu}(k_{\nu})=
\frac{e \eta \gamma}{(2 \pi)^{3}}
\overline{u}_{2} \left[
\frac{1}{M}\sigma_{\mu \nu}k_{\mu}G^{(d)}(\Delta m^2)+
\frac{k^{2}_{\lambda}P_{\mu}-
(k_{\nu}P_{\nu})k_{\mu}}{M^3(k^{2}_{\lambda}-\Delta m^{2})}
[G^{(d)}(k^{2}_{\lambda})- \right. \nonumber \\
 \left. -G^{(d)}(\Delta m^2)]+
 \frac{i}{M^{2}}\epsilon_{\mu \nu \lambda \sigma}P_{\nu}k_{\lambda}
\gamma_{\sigma}\gamma_{5}G^{(T)}(k^{2}_{\lambda})
\right] \gamma_{5}u_{1},
\label{JB}
\end{eqnarray}
where the kinematic notations are the same as in Sect.3.

Remark that the form factors introduced by Eq.(\ref{JA}) do not correspond
to the well-  
defined multipole expansion of currents \cite{Chesh}-\cite{Sachs}.  
That is why we would like to base
our discussion on the Eq.(\ref{JB}) which,
as has been shown explicitly in \cite{BDK},
does correspond to a definite multipole expansion in a properly chosen   
reference system, where $k^{2}_{\mu}=\Delta m^{2}- {\mbox{\boldmath $k$}}^2$.
In this reference system the nonrelativistic reduction of Eq.(\ref{JB})
has the
forms~\cite{BDK}
\begin{itemize}
\item
for electric contribution
\begin{eqnarray}
G^{(d)}(k_{\mu}^{2})\overline{u}_{2}i\sigma_{\mu \nu}k_{\nu}
\gamma_{5}u_{1}A_{\mu}
 &\stackrel{e.k.e.}{\Longrightarrow} &
G^{(d)} (\Delta m^{2} - {\mbox{\boldmath $k$}}^2)
\varphi_{2}^{+} \mbox{\boldmath $\sigma$} \varphi_{1}
[ i\mbox{\boldmath $k$}\Phi  + i \Delta m \mbox{\boldmath $A$} ]
 \nonumber \\
&\Longrightarrow&  G^{(d)} (\Delta m^{2} - {\mbox{\boldmath $k$}}^2)
\varphi_{2}^{+} \mbox{\boldmath $\sigma$} \varphi_{1}
[\mbox{\boldmath $\nabla$} \Phi + \dot{\mbox{\boldmath $A$}}]
 \nonumber \\
&\stackrel{|\mbox{\boldmath $k$}| \to 0}{\Longrightarrow}&
-G^{(d)} (\Delta m^{2})
\varphi_{2}^{+} \mbox{\boldmath $\sigma$} \varphi_{1}
\mbox{\boldmath $E$}
 \nonumber \\
&=:& - \mbox{\boldmath $d$} \mbox{\boldmath $E$},
\label{Gd}
\end{eqnarray}
\item
for toroid contribution
\begin{eqnarray}
-G^{(T)}(k_{\mu}^{2})\overline{u}_{2}i \epsilon_{\mu \nu \rho \lambda}
\gamma_{\nu}
P_{\rho}k_{\lambda} u_{1} A_{\mu}
 &\stackrel{e.k.e.}{\Longrightarrow} &
- G^{(T)}(\Delta m^{2} - {\mbox{\boldmath $k$}}^2)
\mbox{\boldmath $k$} \times [ \mbox{\boldmath $k$} \times
\mbox{\boldmath $\sigma$} ] \mbox{\boldmath $A$} \nonumber \\
&\Longrightarrow& G^{(T)} (\Delta m^{2} - {\mbox{\boldmath $k$}}^2)
\varphi_{2}^{+} \mbox{\boldmath $\sigma$} \varphi_{1} 
\mbox{\boldmath $\nabla$}
\times \mbox{\boldmath $\nabla$} \times \mbox{\boldmath $A$}
 \nonumber \\
&\stackrel{|\mbox{\boldmath $k$}| \to 0}{\Longrightarrow}&
G^{(T)} (\Delta m^{2})\varphi_{2}^{+} \mbox{\boldmath $\sigma$}
\varphi_{1} \mbox{\boldmath $\nabla$}
\times \mbox{\boldmath $B$}
\nonumber  \\
&=:&\mbox{\boldmath $T$} \, \, \mbox{\rm rot} \mbox{\boldmath $B$}.
\end{eqnarray}
\end{itemize}
Here $\mbox{\boldmath $d$}$ and $\mbox{\boldmath $T$}$
are the electric and toroid dipole moments~\cite{Chesh,DT}.

One can see that indeed the parametrization given by the Eq.(\ref{JB}) is
a multipole one where the projections of electric
and toroid transition dipole moments are given, respectively, by
\begin{eqnarray}
d=(e/M)G^{(d)}(\Delta m^{2}) ,
\label{2.81} \\
T=(e/M^{2})G^{(T)}(\Delta m^{2}) .
\label{2.83}
\end{eqnarray}
We remind that the toroid dipole moment transition violates parity but not
T-invariance while the electrical dipole violates both P- and
T-invariance.
The derivatives of formfactors $ G^{(d)}(k^{2}_{\lambda})$ and
$G^{(T)}(k^{2}_{\lambda})$ define the corresponding transition
averaged radii. Since
\begin{equation}
G_{2}(k^{2}_{\lambda})=
G^{(d)}(\Delta m^{2})+\frac{k^{2}_{\lambda}-\Delta m^2}{M \Delta m}
G^{(T)}(k^{2}_{\lambda})
\end{equation}
we obtain approximately~\cite{Chesh,DT}
\begin{equation}
(e/M) G_{2}(\Delta m^2)=d-\Delta m T .
\end{equation}
Note that $\Delta m$ here has pure kinematical origin, that is
with $\Delta m=0$ the decay discussed would not go. Hence, we are forced
conserving the possibility of baryon decays to go over to the threshold
value of $k_{\lambda}=\Delta m^2$ instead of the static point
$k_{\lambda}^{2} = {\mbox{\boldmath $k$}}^{2} = 0$ where
the diagonal moments are usually determined.

The last formula partly resolves a puzzle with the Hara theorem. Indeed in
the $SU(3)_{f}$ limit:
\begin{itemize}
\item The dipole transition moments of the charged hyperon decays
should vanish and presumably stay small due to Ademollo-Gatto theorem
\cite{AG} even in the presence of the $SU(3)_{f}$ breaking terms;
\item The toroid transition dipole moments defined by the Eq.(\ref{2.83})
need not to be zero for these decays as their contributions decouples
automatically in the limit $\Delta m=0 $.
\end{itemize}
So the toroid  transition dipole moment of the
$ \Sigma^{+}\rightarrow p+\gamma $ may be in the origin of the large
asymmetry
observed \cite{PDG}.

\section{The new version of the Hara theorem}
\setcounter{equation}{00}

$\quad$ In order to state our result in another way we write the PV part of the 
weak radiative transition matrix element  with the Lorentz structure 
$O^{T}_{\mu} = i \epsilon_{\mu \nu \lambda \rho}P_{\nu} k_{\lambda} 
\gamma_{\rho}$
in the framework of the  $SU(3)_{f}$ symmetry approach following
strictly \cite{Hara} as
\begin{eqnarray}
M=J^{(T)}_{\mu} \epsilon_{\mu} + H.C.
=\{a^{T}(\overline{B}^{2}_{3}O^{T}_{\mu}B^{1}_{1}+
\overline{B}^{3}_{2}O^{T}_{\mu}B^{1}_{1}
+\overline{B}^{1}_{1}O^{T}_{\mu}B^{2}_{3}+
\overline{B}^{1}_{1}O^{T}_{\mu}B^{3}_{2})+
\nonumber \\
b^{T}(\overline{B}^{3}_{1}O^{T}_{\mu}B^{1}_{2}+
\overline{B}^{2}_{1}O^{T}_{\mu}B^{1}_{3})+
c^{T}(\overline{B}^{1}_{2}O^{T}_{\mu}B^{3}_{1}+
\overline{B}^{1}_{3}O^{T}_{\mu}B^{2}_{1})
\} \epsilon_{\mu},
\label{new}
\end{eqnarray}
where $B_{\alpha}^{\beta}$ is the $SU(3_{f})$ baryon octet, $B_{1}^{3} = p,
\, \, B_{1}^{2} = \Sigma^{+}$ etc., and $a^{T}$ and $b^{T},c^{T}$ are up to 
a factor the toroid dipole moments of the neutral and charged hyperon weak
radiative transitions, respectively.
It is easy to see that this matrix element is invariant under the exchange of
the indices  2 to 3 and { \it vice versa} as it should be in the standard model
of weak interaction. Positive signs in front of every baryon 
bilinear combination
arrive due to Hermitian conjugation properties of the relevant Lorentz
structure $ O^{T}_{\mu}$. Now in Eq.(\ref{new}) all $6$ PV radiative 
transitions are open in contrast to the Hara expression \cite{Hara}:
\begin{eqnarray}
M=J^{(d)}_{\mu} \epsilon_{\mu} + H.C.
= a^{d}(\overline{B}^{2}_{3}O^{d}_{\mu}B^{1}_{1}+
\overline{B}^{3}_{2}O^{d}_{\mu}B^{1}_{1}
-\overline{B}^{1}_{1}O^{d}_{\mu}B^{2}_{3}-
\overline{B}^{1}_{1}O^{d}_{\mu}B^{3}_{2})\epsilon_{\mu},
\label{hara}
\end{eqnarray}
based on another Lorentz structure form $ O^{d}_{\mu}=
i \sigma_{\mu \nu} k_{\nu}\gamma_{5} $ \cite{Hara} 
Due to Hermitian conjugation properties of it 
PV transitions for the
decays $ \Sigma^{+} \rightarrow p+\gamma$ and $ \Xi^{-}\rightarrow\Sigma^{-}
+\gamma $ do not appear in the Eq.(\ref{hara}) as the relevant terms 
$$ \overline{B}^{3}_{1}O^{d}_{\mu}B^{1}_{2}+\overline{B}^{2}_{1}
O^{d}_{\mu}B^{1}_{3},$$ being invariant under exchange of indices
2 to 3 and { \it vice versa} , change sign under Hermitian conjugation 
and so sum up to zero. This is in fact a source of all the troubles with 
the hyperon weak radiative decays evaluations. 
Indeed contributions of the neutral 
hyperon decays $(\Sigma^{0},\Lambda) \Rightarrow n + \gamma $,
$ \Xi^{0} \rightarrow (\Sigma^{0},\Lambda) + \gamma $
and those charged $ \Sigma^{+} \rightarrow p+\gamma$ and 
$ \Xi^{-}\rightarrow\Sigma^{-}+\gamma $
being decoupled, it is difficult to hope that $ SU(3)_{f}$ symmetry breaking
terms would be of the same strength as the coupling constant $ a^{d}$, as
the Ademollo-Gatto theorem \cite{AG} forbids here great corrections. Moreover, 
even any  $ SU(3)_{f}$ symmetry corrections are present, they should be 
more or less equal for pairs $(\Sigma^{0},\Lambda) \Rightarrow n + \gamma $,
$ \Sigma^{+} \rightarrow p+\gamma$
and $ \Xi_{0} \rightarrow \Sigma^{0} + \gamma $,
$ \Xi^{-}\rightarrow\Sigma^{-}+\gamma $
as in both cases there are practically the same differences of masses either 
in charged or in neutral hyperon weak radiative decays. And then one predicts
zero or small asymmetry for the decays
$ \Sigma^{+} \rightarrow p+\gamma$ and 
$ \Xi^{-}\rightarrow\Sigma^{-}+\gamma $
and non-zero (and eventually large) asymmetry for the decays
$(\Sigma^{0},\Lambda) \Rightarrow n + \gamma $ and
$ \Xi^{0} \rightarrow (\Sigma^{0},\Lambda) + \gamma $. As we know 
experimental data give large and negative asymmetry for the decay
$ \Sigma^{+} \rightarrow p+\gamma$ and indicate an asymmetry of the
same order of magnitude for other measurable decays in net contradiction
with the Hara theorem prediction.
Instead the parameters $b^{T},c^{T}\sim T$ 
in Eq.(\ref{new})  opens
a possibility to account for large nonzero asymmetry in the charged
hyperon radiative decays even in the $SU(3)_{f}$ symmetry limit and
thus eventually overcome a contradiction with experiment.
We display in the Table 1 the experimental data from \cite{PDG} and
in the Table 2 the results of Eq.(\ref{new}) (see the 3rd column of the 
Table 2) and \cite{Hara} (see the 2nd column of the Table 2). 
We have also put there  results of a traditional 
single-quark radiative transition which we have taken from \cite{Sharma} 
just to show in what way our new current in Eq.(\ref{new}) 
can reproduce results of a quark model
in a single-quark diagram approximation (that is with a single quark
weak radiative transition $ s \rightarrow d + \gamma  $, 
two other quarks being spectators). For that purpose it is sufficient to
assume that hyperon radiative decays in the $SU(3)_{f}$ model are described
by the effective $|\Delta S=1|$ octet neutral weak current similar in
form to that of the effective $|\Delta S=1|$ nonleptonic hyperon transitions
(see, e.g.,~\cite{Marshak}):
\begin{eqnarray}
J_{\mu}^{W}=(-F+D)\overline{B}^{3}_{\gamma}O^{T}_{\mu}B^{\gamma}_{2}+
(F+D)\overline{B}^{\gamma}_{2}O^{T}_{\mu}B^{3}_{\gamma}.
\label{wnc}
\end{eqnarray}
This {\it ad hoc} assumption in the usual $SU(6)$ symmetry
limit ( that is with $F=2/3D$) which corresponds to use of the 
nonrelativistic quark model,
yields the known results of the single-quark approach as can be easily seen
from the Table 2 (note a sign difference for the $\Xi$ hyperon wave function in
\cite{Sharma} while putting $F=-2/3b,D=-b$).
Instead with the $F=0$ one arrives at the results 
given by the new formulation of the Hara theorem with $-a^{T}=b^{T}=c^{T}=D$.
This indicates explicitly a source of contradiction between the Hara theorem
\cite{Hara} and single-quark model \cite{Sharma} predictions. 

Assuming that all asymmetry in the decay $ \Sigma^{+} \rightarrow p+\gamma$ 
is given by the toroid dipole transition moment we have found an upper
bound for it from the experimental data \cite{PDG} using 
Eqs.(\ref{alpha}, \ref{rate}, \ref{new}) to be
$ |T| < 10^{-34}$ (cm$^{2}$). This value turns out to be close to 
the predicted values
of the toroid dipole moments of the neutrinos $\nu_{\mu}$
($ T_{\nu_{\mu}}\approx e[+1.090 \, \, \mbox{to} \, \,  +2.329] \times 10^{-34}$ (cm$^{2}$)) 
and $\nu_{\tau}$
($ T_{\nu_{\tau}} \approx e[-1.971 \, \, \mbox{to} \, \, -0.732] \times 10^{-34}$ (cm$^{2}$))
but noticeably lower than that of the $\nu_{e}$ neutrino
($ T_{\nu_{e}} \approx e[+6.873 \, \, \mbox{to} \, \, 8.112] \times 10^{-34}$ (cm$^{2}$)) \cite{Valya}.


\section{New derivation of the Vasanti formula }
\setcounter{equation}{00}


$\quad$ Radiative hyperon decays were analyzed first in \cite{Vasa}
at the quark level upon taking into account
chiral invariance considerations. Later similar results were obtained
with the Feynman diagram technique in the frame of the Weinberg-Salam model
\cite{Gad}.
We shall try to re-derive 
the main result of \cite{Vasa}, namely, that the PV single-quark weak
radiative transition $ s \rightarrow d + \gamma $ is 
proportional to $ (m_{s}-m_{d}) $.
Similar to \cite{Vasa} we assume that quarks are on their mass shell.
So at the quark level we write  instead of the Vasanti formula
$$ M=\frac{Ge}{\sqrt{2}}sin\theta cos \theta
\overline{d}\left[a+b\gamma_{5}\right]
i \sigma_{\mu \nu} k_{\nu} \epsilon_{\mu}s $$
for the amplitude of the $ s \rightarrow d+ \gamma $ decay
with $ G, e, \theta $ being Fermi constant, 
unit of charge and Cabbibo angle, respectively, 
$ q (q=d,s)$ also meaning here spinor 
of the quark $ q$ with the momentum $ p_{q}$,
another one, using the Lorentz structure $ O^{T}_{\mu}=
i \epsilon_{\mu \nu \lambda \rho}P_{\nu} k_{\lambda} \gamma_{\rho}, $
where now $ P_{\nu}=(p_{s}+p_{d})_{\nu}, \, k_{\nu}=(p_{s}-p_{d})_{\nu}$:
\begin{equation}
M=\frac{Ge}{\sqrt{2}}sin\theta cos \theta
\overline{d} \gamma_{5}(a'+b' \gamma_{5})i \epsilon_{\mu \nu \lambda \rho}
P_{\nu} k_{\lambda} \gamma_{\rho} \epsilon_{\mu} \gamma_{5} s,
\label{vasa}
\end{equation}
and upon using generalized Gordon identities,
where now all quark quantities are
assumed ( quarks are on their mass shell), arrive at
\begin{equation}
M=\frac{Ge}{\sqrt{2}}sin\theta cos \theta
\overline{d}\left[a'(m_{s}+m_{d})+b'(m_{s}-m_{d})\gamma_{5}\right]
i \sigma_{\mu \nu} k_{\nu} \epsilon_{\mu}s,
\label{M}
\end{equation}
that is, in fact the main Vasanti result \cite{Vasa} is reproduced.
The factors $(m_{s}\pm m_{d})$ arrive due to the generalized Gordon
identities. The relative signs of $a'$ and $b'$ are not fixed here so
it is possible to obtain negative value of the asymmetry parameter. 
Note that Eq.(\ref{M})
(with $a'=b'=1$) was obtained in \cite{Vasa} upon assuming (i) chiral
invariance, (ii) validity of the original Hara theorem. We have proved
in fact that the introduction of the Lorentz structure $O^{T}$ 
at the quark level
is equivalent to the chiral invariance approach of \cite{Vasa} and to
the diagram calculation result of \cite{Gad}. This result dictates
the insertion of the factor $(m_{s}-m_{d})$ into the parameter $c$ 
(see the 2nd column of the Table 2 and single-quark transition terms in
\cite{Sharma,Verma} and other works cited in \cite{ZenM}) to assure
the correct behavior of the corresponding quark PV transition amplitudes.
And { \it vice versa}, the results of \cite{Vasa} and \cite{Gad} together
with the generalized Gordon identities show that at the 
quark level it is a toroid dipole moment  with 
its characteristic Lorentz structure  
$O^{T}=i\epsilon_{\mu \nu \lambda \rho}P_{\nu}k_{\lambda} \gamma_{\rho}$ 
which is generated  through the Feynman diagrams contributions of \cite{Gad}.
This result is valid also for the penguin diagram contributions (see, e.g.,
\cite{Penguin}) which have similar to Eq.(\ref{M}) Lorentz structure 
for the case of a real photon emission. Indeed, single-quark 
contributions yield the values of the toroid dipole moment $T$
for the  $ s \rightarrow d + \gamma $
decay at the level of $\sim e 10^{-35}$cm$^{2}$, while QCD one-loop corrections
(for recent calculations see, e.g., \cite{Fabbri}) have a trend to 
diminish it drastically. The long-distance corrections (see, e.g.
\cite{Mendel}) could raise it by an order of magnitude.

\section{Two-quark weak radiative transitions }
\setcounter{equation}{00}

It is known that single-quark transitions $ s \rightarrow d + \gamma $
give only a small part of the hyperon weak radiative decay rate.
It can be seen already from the analysis of experimental data on $ \Sigma^{+} 
 \rightarrow p \gamma $ and $ \Xi^{-} \rightarrow \Sigma^{-} \gamma$.
The decay  $ \Xi^{-} \rightarrow \Sigma^{-} \gamma$ can go only through
a single-quark diagram and its branching ratio occurs to be at 
the level of $ 10^{-4}$, while other 
measured decays, including the decay $ \Sigma^{+} \rightarrow p \gamma $
have branching ratios an order of magnitude higher,  $ 10^{-3}$.
So although we have
shown that a single-quark transition $ s \rightarrow d + \gamma $
can be understood in terms of
the toroid dipole moment it proves to be inadequate alone to describe
experimental data in the framework of a quark model. Neither penguin
diagram contributions are strong enough to enhance 
an effective $ s \rightarrow d+ \gamma $ \cite{Penguin}. This result is
not surprising as the same has been proved to be true for the
effective $ q \rightarrow q'+\gamma$ where $q,q'$ are light quarks
\cite{Donoghue,DZZ}.
So one has  also to consider the
two-quark weak radiative transitions $ s+u \rightarrow u+d + \gamma $ 
which proceed via W-exchange and appears to be dominant.
It has been done in the thoughtful works of \cite{Sharma,Verma}
and others and compilated in a very complete review of \cite{Zen}.
In the calculations it has been assumed that all the external quarks
are on their mass shell. The decay amplitudes were evaluated by sandwiching 
two-quark weak radiative transition operator  between the baryon wave
functions, one of the quarks being a spectator. 
We show here that the results of \cite{Sharma,Verma}
can be obtained within the nonrelativistic quark model (NRQM).
In this way a treatment of the hyperon weak radiative decays 
comes close to those of magnetic moments and
weak $ \beta$-decay coupling constants in the framework of NRQM 
\cite{Morp},\cite{Gas}. The results of \cite{Sharma} do not 
rely heavily
on the choice of the current quarks.  Indeed, if one wants to treat
the quark diagrams properly one should use current quarks. But the
internal symmetry of two-quark $W$-exchange contributions into 
the hyperon radiative decays can be understood, as we shall see,
already at the level of the nonrelativistic quark model with the 
baryon wave functions given by the $ SU(6)$ model.
Let us just start from the NRQM diagrams of the kind given in Fig.1 and Fig.2. 
We shall write here at some length the way of our reasoning for 
$ \Sigma^{+} \rightarrow p \gamma $ decay, putting other decays into 
Appendix A. 
Thus the $ \Sigma^{+} \rightarrow p \gamma $ decay amplitude can be 
casted in the form
\begin{eqnarray}
6<p_{\downarrow}, \gamma(+1) |O| \Sigma^{+}_{\uparrow} >=\nonumber\\
<2u_{2}u_{2}d_{1}-u_{2}d_{2}u_{1}-
d_{2}u_{2}u_{1},\gamma(+1)|O|2u_{1}u_{1}s_{2}-u_{1}s_{1}u_{2}-
s_{1}u_{1}u_{2}>=\nonumber\\
4<u_{2}u_{2}d_{1},\gamma(+1) |O|u_{1}u_{1}s_{2}>-
4<u_{2}u_{2}d_{1},\gamma(+1) |O|u_{1}s_{1}u_{2}>- \nonumber\\
4<u_{2}d_{2}u_{1},\gamma(+1) |O|u_{1}u_{1}s_{2}>+
4<u_{2}d_{2}u_{1},\gamma(+1) |O|u_{1}s_{1}u_{2}>,
\label{sigma}
\end{eqnarray}
where $ q_{1,2}$ just mean the helicity state $ q_{\uparrow,\downarrow} $ 
of the quarks inside the baryon, respectively. The $O$ is an operator
which we do not need to write explicitly here.
The 1st matrix element on the RHS of the Eq.(\ref{sigma})
$ <u_{2}u_{2}d_{1},\gamma(+1) |O|u_{1}u_{1}s_{2}>$
in the case of W-exchange between the quarks can be described
only by the diagram Fig.1(1), as there is not possible to 
represent it  by a diagram with a spectator quark. We disregard it following
the reasons of \cite{Verma}. Really it is plausible that three-quark transition
involving $W$-exchange between two quarks and a photon emission by the 3rd
quark is suppressed due to kinematical reasons. Instead
the 2nd matrix element on the RHS of the Eq.(\ref{sigma})
$ <u_{2}u_{2}d_{1},\gamma(+1) |O|u_{1}s_{1}u_{2}>=A_{1} $
can be described by three different diagrams Fig.2(1) with the
quark $ q_{2}$ as a spectator, 
$$ A_{1}=\frac{2}{3}A-\frac{1}{3}E+\frac{2}{3}B. $$
Here $A$ corresponds to the helicity non-flip weak transition amplitude
with all quark heaving helicity `up',
 $ s_{\uparrow}+u_{\uparrow} \rightarrow u_{\uparrow}+d_{\uparrow} $,
$E$ corresponds to the helicity non-flip weak transition amplitude 
with different helicities of the interacting quarks
 $ s_{\downarrow}+u_{\uparrow} \rightarrow u_{\downarrow}+d_{\uparrow} $
while $B$ corresponds to the helicity-flip weak transition amplitude
with the `up' helicity of the $ s$ quark 
 $ s_{\uparrow}+u_{\downarrow} \rightarrow u_{\downarrow}+d_{\uparrow} $,
the factors 2/3 and -1/3 are just the values of the quark charges in term
of the proton electric charge $e$.
The 3nd matrix element on the RHS of the Eq.(\ref{sigma})
$ <u_{2}d_{2}u_{1},\gamma(+1) |O|u_{1}u_{1}s_{2}>=A_{3}$
can be described by three different diagrams Fig.2(3) with the
quark $ q_{1}$ as a spectator, 
$$ A_{3}=\frac{2}{3}C-\frac{1}{3}E+\frac{2}{3} \tilde A, $$
with two new quantities,
$C$ corresponding to the helicity-flip weak transition amplitude
with the `down' helicity of the $ s$ quark,
 $ s_{\downarrow}+u_{\uparrow} \rightarrow u_{\uparrow}+d_{\downarrow} $, 
and $ \tilde A$ corresponding to the helicity non-flip weak transition 
amplitude with down helicities of the interacting quarks, 
 $ s_{\downarrow}+u_{\downarrow} \rightarrow u_{\downarrow}+d_{\downarrow}. $
The 4nd matrix element on the RHS of the Eq.(\ref{sigma})
$ <u_{2}d_{2}u_{1},\gamma(+1) |O|u_{1}s_{1}u_{2}>$
is described by two sets of diagrams, these given by the Fig.2(2)
with quark $ q_{2}$ as a spectator, 
$$ A_{2}=-\frac{1}{3}A-\frac{1}{3}C+\frac{2}{3}D, $$
where a new quantity
$D$ corresponds to the helicity non-flip weak transition amplitude 
with different helicities of the interacting quarks, 
 $ s_{\uparrow}+u_{\downarrow} \rightarrow u_{\uparrow}+d_{\downarrow} $,
and those given by the Fig.2(4)
with the quark $ q_{1}$ as a spectator, 
$$ A_{4}=\frac{2}{3}D-\frac{1}{3} \tilde A-\frac{1}{3}B, $$

So finally
\begin{equation}
<p_{\downarrow}, \gamma(+1) |O| \Sigma^{+}_{\uparrow} >=
\frac{2}{3}(-2A_{1}+A_{2}-2A_{3}+A_{4})
\end{equation}

If we assume that a spectator quark does not induce changes in the $ A_{k},
\, k=1, 2, 3, 4,$
in this approximation all the hyperon radiative decays can be written in terms 
of the quantities $ A_{1,2,3,4} $: 
\begin{eqnarray}
<p_{\downarrow}, \gamma(+1) |O| \Sigma^{+}_{\uparrow} >=
\frac{2}{3}(-2A_{1}+A_{2}-2A_{3}+A_{4}),
\nonumber\\
<n_{\downarrow}, \gamma(+1) |O| \Sigma^{0}_{\uparrow} >=
\frac{2}{3\sqrt{2}}(A_{1}-2A_{2}-2A_{3}+A_{4}),
\nonumber\\
<n_{\downarrow}, \gamma(+1) |O| \Lambda_{\uparrow} >=
\frac{2}{\sqrt{6}}(A_{1}-2A_{2}-A_{4}),
\nonumber\\
<\Lambda_{\downarrow}, \gamma(+1) |O| \Xi^{0}_{\uparrow} >=
\frac{2}{\sqrt{6}}(A_{1}-A_{2}),
\nonumber\\
<\Sigma^{0}_{\downarrow}, \gamma(+1) |O| \Xi^{0}_{\uparrow} >=
\frac{2}{3 \sqrt{2}}(A_{1}+A_{2}-2A_{3}+4A_{4}).
\label{AAA}
\end{eqnarray}
Two-quark $W$-exchange amplitudes satisfy two relations:
$$
<p_{\downarrow}, \gamma(+1) |O| \Sigma^{+}_{\uparrow} >+
2\sqrt{6}<\Lambda_{\downarrow}, \gamma(+1) |O| \Xi^{0}_{\uparrow} >=
$$
$$ \sqrt{2}<\Sigma^{0}_{\downarrow}, \gamma(+1) |O| \Xi^{0}_{\uparrow} >+
\sqrt{6}<n_{\downarrow}, \gamma(+1) |O| \Lambda_{\uparrow} >,
$$
$$
\sqrt{2}<n_{\downarrow}, \gamma(+1) |O| \Sigma^{0}_{\uparrow} >=
<p_{\downarrow}, \gamma(+1) |O| \Sigma^{+}_{\uparrow} >+
\sqrt{6}<\Lambda_{\downarrow}, \gamma(+1) |O| \Xi^{0}_{\uparrow} >,
$$
as they depend not on all $ A_{k}$`s, but only on their linear combinations
$ A_{1}- A_{2},\, A_{2}+ A_{3},\, A_{3}- A_{4}$.
It is straightforward to show that with
\begin{eqnarray}
A_{1}^{PV}=\frac{1}{6}+\frac{2}{3}X-\zeta(\frac{1}{3}+\frac{4}{3}X),\quad
A_{2}^{PV}=-\frac{1}{6}+\frac{2}{3}X-\zeta (-\frac{1}{3}+\frac{4}{3}X),
\nonumber\\
A_{3}^{PV}=\frac{1}{6}-\frac{1}{3}X-\zeta X,\quad
A_{4}^{PV}=-\frac{1}{6}-\frac{2}{3}X-\zeta X,\nonumber\\
A_{1}^{PC}=\frac{1}{6}+\frac{2}{3}X-\zeta(\frac{1}{3}+\frac{4}{3}X),\quad
A_{2}^{PC}=-\frac{1}{6}+\frac{2}{3}X-\zeta (-\frac{1}{3}+\frac{4}{3}X),
\nonumber\\
A_{3}^{PC}=-\frac{1}{6}-\frac{2}{3}X-\zeta X,\quad
A_{4}^{PC}=\frac{1}{6}-\frac{1}{3}X-\zeta X,
\end{eqnarray}
where $ X=k/2m_{u}$ and $ 6\zeta=(1-m_{u}/m_{s}) $ \cite{Sharma},
one arrives  up to  an overall constant exactly at the results of 
\cite{Sharma} (See Table 1 in \cite{Sharma} and the 3rd column of the Table 3 
of this work) and  with $ X=0,\, \epsilon=1-6\zeta=0$ 
we return to the results of \cite{Zen}.

So the main results of the two-quark weak radiative transition model can be 
understood in the framework of NRQM in terms of the quantities   
$ A_{k}, \, k=1,2,3,4,$, which however cannot be calculated without 
further assumptions. Neither we have answered the question where is the 
source of a contradiction between the two-quark transition model results 
and those of Hara \cite{Hara}. In order to answer at least partly to 
this question we shall proceed in a way 
similar to that of the \cite{Sharma} and \cite{China}, returning to 
current quarks within the Salam-Weinberg model.    
Let us consider kinematics of one of the two-quark PV weak radiative 
transition in some detail.
PV part of the Feynman diagram  for the bremsstrahlung process
$ s+u \rightarrow u+d+\gamma $ with the 3rd quark $q$ as a spectator,
where $\gamma$-quantum irradiates from the $u$-quark,
reads \cite{Sharma} (we write
$ q(p_{k}), q=u, d, s, k=1, \ldots 6,$ for a spinor of a quark $q$ with momentum 
$p_{k}$, while $(ab)$ means below a scalar product $ a_{\mu}b_{\mu})$:
\begin{eqnarray}
<u(p_{2}),d(p_{4}),q(p_{6}), \gamma(k)|O^{PV}|s(p_{1}),u(p_{3}),q(p_{5})>=
\nonumber\\
e_{u}/(p_{2}k) \overline{u}(p_{2}) \hat k \hat \epsilon(k) 
\gamma_{\mu} \gamma_{5}s(p_{1}) \overline{d}(p_{4}) \gamma_{\mu}u(p_{3})
\overline{q}(p_{6})q(p_{5})+
\nonumber\\
e_{u}/(p_{2}k) \overline{u}(p_{2}) \hat k \hat \epsilon(k) 
\gamma_{\mu}s(p_{1}) \overline{d}(p_{4}) \gamma_{\mu} \gamma_{5}u(p_{3})
\overline{q}(p_{6})q(p_{5}).
\label{due}
\end{eqnarray}
Starting from similar expression, two-quark diagrams were calculated 
in the Coulomb gauge \cite{Sharma}, upon carrying out an expansion of the 
amplitude given by the Fig.1(1) in photon momentum $k$. (Recently
with another technique similar calculations have been performed 
for the bremsstrahlung process
$ b+u \rightarrow u+s+\gamma $ \cite{China}.)
Instead we use an identity (see, e.g., \cite{Gas})
for the 1st term in the RHS of the Eq.(\ref{due})
$$ \gamma_{\alpha}\gamma_{\beta}\gamma_{\rho}=
(\delta_{\alpha \beta}\delta_{\rho \delta}-
\delta_{\alpha \rho}\delta_{\beta \delta}+
\delta_{\alpha \delta}\delta_{\beta \rho})\gamma_{\delta}-
\epsilon_{\alpha \beta \rho \delta}\gamma_{\delta}\gamma_{5} $$
in order to rewrite it ( up to a factor  $ e_{u}/(p_{2}k) $ )
in the form
\begin{eqnarray}
k_{\alpha } \epsilon_{\beta}(k)
\overline{u}(p_{2}) \gamma_{\alpha} \gamma_{\beta}  
\gamma_{\mu} \gamma_{5}s(p_{1})V_{\mu}
\overline{q}(p_{6})q(p_{5}) =
\overline{u}(p_{2})\{
(k \epsilon)\gamma_{\mu}\gamma_{5}s(p_{1})V_{\mu} -\nonumber\\
\hat k\gamma_{5}s(p_{1}) (V \epsilon) +
\hat \epsilon\gamma_{5}s(p_{1}) (V k)-
\epsilon_{\alpha \beta \mu \delta} k_{\alpha } \epsilon_{\beta}(k)
\gamma_{\delta}s(p_{1})V_{\mu}\} \overline{q}(p_{6})q(p_{5}),
\label{my}
\end{eqnarray}
where $$ V_{\mu}=\overline{d}(p_{4})\gamma_{\mu}u(p_{3}).$$
Upon using Gordon identity, assuming the equality of the masses of quarks
$ d$ and $u$ and of their momenta, we obtain that
$$ V_{\mu}=\frac{1}{2m_{u,d}}p_{4\mu}\overline{d}(p_{4})u(p_{3}). $$
that is, the vector $ V_{\mu}$ is
proportional to the momenta of
quarks $ u(p_{3})$ and/or $ d(p_{4}) $. In a 
nonrelativistic quark model
it is reasonable to assume at least for the hyperon decays
$ \Sigma^{+} \rightarrow p \gamma $ and $ (\Sigma^{0},\Lambda)
 \rightarrow n \gamma $ 
that the momenta of the quarks of the final nucleon
are equal, that is $ p_{2}=p_{4}=p_{6}$, and also for the spectator quark the 
equality $ p_{5}=p_{6}$ holds. In this oversimplified picture kinematics of the
hyperon decay is related to the kinematics of the quarks as
$ k=P_{i}-P_{f}=p_{1}-p_{2},\quad P=P_{i}+P_{f}=p_{1}+5p_{2}$
and $ p_{4}=\frac{1}{6}(P-k)$.
The last term in the RHS of the Eq.(\ref{my}) then gives
the structure $ \epsilon_{\alpha \beta \mu \delta} 
k_{\alpha } \epsilon_{\beta} P_{\mu}\gamma_{\delta} $ characteristic for
the toroid dipole transition.
This explains explicitly why the quark model calculations of 
the two-quark contributions into the  weak radiative transition
$ \Sigma^{+} \rightarrow p \gamma $ 
do not follow the Hara theorem. The important part of their contribution
has a Lorentz structure different from that used in the Hara theorem and with
another properties under Hermitian conjugation. An estimation of the second 
term in the RHS of the Eq.(\ref{due}) shows that its contribution
does not change our result. The same is true for other four hyperon weak
radiative decays $(\Sigma^{0},\Lambda) \Rightarrow n + \gamma $ and
$ \Xi^{0} \rightarrow (\Sigma^{0},\Lambda) + \gamma $.
Note that we have obtained 
charged axial-vector current multiplied by the charged scalar current
(and by the neutral scalar current of the spectator quark). But our
conclusion for the PV transition amplitude as a whole remains the same while
Fierz rearrangement
would allow to obtain an expression
similar to that of the single-quark transition given by Eq.(\ref{vasa})
and consequently to arrive at
the description of the radiative transition as a whole in terms of 
hyperons (cf. Eq.(\ref{new})). We shall write it in more detail elsewhere.

\section{Summary and Conclusion}
\setcounter{equation}{00}


$\quad$ In order to resolve a contradiction between the experiments claiming
large
negative asymmetry in $ \Sigma^{+}\rightarrow p+\gamma $, the Hara 
theorem, predicting zero asymmetry for
$ \Sigma^{+}\rightarrow p+\gamma $ and $ \Xi^{-}\rightarrow 
\Sigma^{-}+\gamma $
in the exact  $SU(3)_{f}$ symmetry 
and quark models which cannot reproduce the Hara theorem
results without making vanish all asymmetry parameters in the 
$SU(3)_{f}$ symmetry limit,
we have considered a parity-violating part of the transition
electromagnetic current in the alternative form allowing well-defined multipole
expansion. Part of it which is connected with the Lorentz structure
$i \epsilon_{\mu \nu \lambda \rho}P_{\nu} k_{\lambda} \gamma_{\rho} $
enables as to reformulate the Hara theorem thus opening a possibility
of nonzero asymmetry parameters for all $6$ weak radiative hyperon decays
and revealing hitherto unseen transition toroid dipole moments. 
In this way at least partly it is resolved a long-stayed puzzle
with Hara theorem prediction and experimental result for the asymmetry
of the weak radiative decay $ \Sigma^{+}\rightarrow p+\gamma $. 
We have also reproduced Vasanti formula at the quark level. 
Our result is consistent with the traditional results of the single-quark 
transition models and is unaltered by the QCD corrections including the
penguin diagram contributions.
As to the two-quark weak radiative
transitions we have found that the main part of the
diagram contribution is also connected with the Lorentz structure
$ i\epsilon_{\mu \nu \lambda \rho}P_{\nu}k_{\lambda} \gamma_{\rho}$
which seems to be an intrinsic reason of the apparent contradiction between
the Hara theorem conclusion and quark model results for hyperon weak 
radiative decays.

\section*{Acknowledgments}
\setcounter{equation}{00}

One of the authors (V.Z.) thanks F.~Hussain, N.~Paver and S.~Petcov 
for interest 
to the work and discussion.
One of the authors (V.Z.) is grateful to the International Centre for
Theoretical physics, Trieste, where part of this work has been done, for
hospitality and financial support.

\newpage

$$ { \bf APPENDIX \quad A } $$
$$ { \bf Two-quark \quad W-exchange \quad diagrams \quad
in\quad the \quad NRQM}$$
\begin{enumerate}
\item[(i)]
The decay $  \Sigma^{+} \rightarrow p \gamma$ is described 
in the main text.
\item[(ii)]
The next one is though unseen but important for model reason
the $ \Sigma^{0} \rightarrow n \gamma$ decay.
$$6\sqrt{2}<n_{\downarrow}, \gamma(+1) |O| \Sigma^{0}_{\uparrow}>=$$
$$ <2d_{2}d_{2}u_{1}-d_{2}u_{2}d_{1}-
u_{2}d_{2}d_{1},\gamma(+1)|O|2u_{1}d_{1}s_{2}+2d_{1}u_{1}s_{2}-$$
$$u_{1}s_{1}d_{2}-s_{1}u_{1}d_{2}-d_{1}s_{1}u_{2}-s_{1}d_{1}u_{2}>=$$
$$8<d_{2}d_{2}u_{1},\gamma(+1) |O|u_{1}d_{1}s_{2}>-
8<d_{2}d_{2}u_{1},\gamma(+1) |O|u_{1}s_{1}d_{2}>-\qquad (A.1)$$
$$4<d_{2}d_{2}u_{1},\gamma(+1) |O|d_{1}s_{1}u_{2}>- 
8<d_{2}u_{2}d_{1},\gamma(+1) |O|u_{1}d_{1}s_{2}>+$$
$$4<d_{2}u_{2}d_{1},\gamma(+1) |O|u_{1}s_{1}d_{2}>+
4<d_{2}u_{2}d_{1},\gamma(+1) |O|d_{1}s_{1}u_{2}>$$
The 1st and the 3rd matrix elements on the RHS of the Eq.(A.1)
in the case of W-exchange between the quarks can be described
only by the diagrams Fig.1(2) and Fig.1(3), and we disregard them
following, as its cannot  be
represented by diagrams with a spectator quark.
The 2nd matrix element on the RHS of the Eq.(A.1)
$ <d_{2}d_{2}u_{1}, \gamma(+1) |O|u_{1}s_{1}d_{2}>= A_{2} $
can be described by three diagrams Fig.2(2) with the
quark $ d_{2}$ as a spectator.
The 4nd matrix element on the RHS of the Eq.(A.1)
$ <d_{2}u_{2}d_{1}, \gamma(+1) |O|u_{1}d_{1}s_{2}>= A_{3} $
can be described by three diagrams Fig.2(3) with the
quark $ d_{1}$ as a spectator.
The 5th matrix element on the RHS of the Eq.(A.1) is
$ <d_{2}u_{2}d_{1}, \gamma(+1) |O|u_{1}s_{1}d_{2}>= A_{1} $ given by
three diagrams of the Fig.2(1) with the
quark $ d_{2}$ as a spectator.
The 6th matrix element on the RHS of the Eq.(A.1) is
$ <d_{2}u_{2}d_{1},\gamma(+1) |O|d_{1}s_{1}u_{2}>=A_{4} $ given by
three diagrams of the Fig.2(4)  with the
quark $ d_{1}$ as a spectator.
So
$$ <n_{\downarrow}, \gamma(+1) |O| \Sigma^{0}_{\uparrow} >=
\frac{2}{3\sqrt{2}}(A_{1}-2A_{2}-2A_{3}+A_{4}).$$
\item[(iii)]
The $ \Lambda \rightarrow n \gamma $ decay amplitude can be
written in the form
$$ 2 \sqrt{6}<n_{\downarrow}, \gamma(+1) |O| \Lambda_{\uparrow} >=$$
$$ <2d_{2}d_{2}u_{1}-d_{2}u_{2}d_{1}-u_{2}d_{2}d_{1},\gamma(+1)|O|
u_{1}s_{1}d_{2}+s_{1}u_{1}d_{2}-d_{1}s_{1}u_{2}-s_{1}d_{1}u_{2}>=$$
$$ 4<d_{2}d_{2}u_{1},\gamma(+1)|O|u_{1}s_{1}d_{2}>-
4<d_{2}d_{2}u_{1},\gamma(+1)|O|d_{1}s_{1}u_{2}>-\qquad (A.2)$$
$$ 4<d_{2}u_{2}d_{1},\gamma(+1)|O|u_{1}s_{1}d_{2}>+
4<u_{2}d_{2}d_{1},\gamma(+1)|O|d_{1}s_{1}u_{2}>.$$
The 1st matrix element on the RHS of the Eq.(A.2)
$ <d_{2}d_{2}u_{1},\gamma(+1) |O|u_{1}s_{1}d_{2}>=A_{2} $
can be described by three diagrams Fig.2(2) with the
quark $ d_{2}$ as a spectator.
The 2nd matrix element on the RHS of the Eq.(A.2)
$ <d_{2}d_{2}u_{1},\gamma(+1) |O|d_{1}s_{1}u_{2}>$
in the case of W-exchange between the quarks can be described
only by the diagram Fig.1(3), as it cannot  be
represented by a diagram with a spectator quark.
The 3rd matrix element on the RHS of the Eq.(A.2) 
$ <d_{2}u_{2}d_{1},\gamma(+1) |O|u_{1}s_{1}d_{2}>=A_{1} $
can be described by three diagrams Fig.2(1) with the
quark $ d_{2}$ as a spectator.
The 4th matrix element on the RHS of the Eq.(A.2)
$ <d_{2}u_{2}d_{1},\gamma(+1) |O|d_{1}s_{1}u_{2}>=A_{4} $
can be described by three diagrams Fig.2(4) with the
quark $ d_{1}$ as a spectator. So
$$ <n_{\downarrow}, \gamma(+1) |O| \Lambda_{\uparrow} >=
\frac{2}{\sqrt{6}}(A_{1}-2A_{2}-A_{4}).$$
\item[(iv)] 
The $ \Xi^{0} \rightarrow \Lambda \gamma $ decay amplitude can be
written in the form
$$ 2 \sqrt{6}<\Lambda_{\downarrow}, \gamma(+1) |O| \Xi^{0}_{\uparrow} >=$$
$$ <u_{2}s_{2}d_{1}+s_{2}u_{2}d_{1}-d_{2}s_{2}u_{1}-s_{2}d_{2}u_{1},
\gamma(+1)|O|
2s_{1}s_{1}u_{2}-s_{1}u_{1}s_{2}-u_{1}s_{1}s_{2}>=$$
$$ 4<u_{2}s_{2}d_{1}, \gamma(+1) |O|s_{1}s_{1}u_{2}>-
4<u_{2}s_{2}d_{1}, \gamma(+1) |O|s_{1}u_{1}s_{2}>-\qquad (A.3)$$
$$ 4<d_{2}s_{2}u_{1}, \gamma(+1) |O|s_{1}s_{1}u_{2}>+
4<d_{2}s_{2}u_{1}, \gamma(+1) |O|s_{1}u_{1}s_{2}>.$$
The 1st and 3rd matrix elements on the RHS of Eq.(A.3)
in the case of W-exchange between the quarks can be described
only by the diagrams Fig.1(4) and Fig.1(5), and we disregard them,
as they cannot  be
represented by diagrams with a spectator quark.
The 2nd matrix element on the RHS of the Eq.(A.3)
$ <u_{2}s_{2}d_{1}, \gamma(+1) |O|s_{1}u_{1}s_{2}>= A_{1}$ 
can be described by three diagrams Fig.2(1) with the
quark $ s_{2}$ as a spectator.
The 4th matrix element on the RHS of the Eq.(A.3)
$<d{2}s_{2}u_{1}, \gamma(+1) |O|s_{1}u_{1}s_{2}>= A_{2}$ 
can be described by three diagrams Fig.2(2) with the
quark $ s_{2}$ as a spectator. So
$$ <\Lambda_{\downarrow}, \gamma(+1) |O| \Xi^{0}_{\uparrow} >=
\frac{2}{\sqrt{6}}(A_{1}-A_{2}).$$
\item[(v)] 
The $ \Xi^{0} \rightarrow \Lambda \gamma $ decay amplitude can be 
written in the form
$$ 6 \sqrt{2}< \Sigma^{0}_{ \downarrow}, \gamma(+1) |O| \Xi^{0}_{\uparrow}>=
<2u_{2}d_{2}s_{1}+2d_{2}u_{2}s_{1}$$
$$ -u_{2}s_{2}d_{1}-s_{2}u_{2}d_{1}-d_{2}s_{2}u_{1}-s_{2}d_{2}u_{1},
\gamma(+1)|O|
2s_{1}s_{1}u_{2}-s_{1}u_{1}s_{2}-u_{1}s_{1}s_{2}>=$$
$$8<u_{2}d_{2}s_{1}, \gamma(+1) |O|s_{1}s_{1}u_{2}>-
8<u_{2}d_{2}s_{1}, \gamma(+1) |O|s_{1}u_{1}s_{2}>-\qquad (A.4)$$
$$ 4<u_{2}s_{2}d_{1}, \gamma(+1) |O|s_{1}s_{1}u_{2}>+
4<u_{2}s_{2}d_{1}, \gamma(+1) |O|s_{1}u_{1}s_{2}>- $$
$$ 4<d_{2}s_{2}u_{1}, \gamma(+1) |O|s_{1}s_{1}u_{2}>+
4<d_{2}s_{2}u_{1}, \gamma(+1) |O|s_{1}u_{1}s_{2}>.$$
The 1st matrix element on the RHS of the Eq.(A.4)
$ <u_{2}d_{2}s_{1}, \gamma(+1) |O|s_{1}s_{1}u_{2}>=A_{4} $  
can be described by three diagrams Fig.2(4) with the   
quark $ s_{1}$ as a spectator.
The 2nd matrix element on the RHS of the Eq.(A.4)
$ <u_{2}d_{2}s_{1}, \gamma(+1) |O|s_{1}u_{1}s_{2}>=A_{3}$
can be described by three diagrams Fig.2(3) with the
quark $ s_{1}$ as a spectator.
The 3rd and 5th matrix elements on the RHS of Eq.(A.4)
in the case of W-exchange between the quarks can be described
only by the diagrams Fig.1(4) and Fig.1(5), and we disregard them,
as they cannot  be
represented by diagrams with a spectator quark.  
The 4th matrix element on the RHS of Eq.(A.4)
$ <u_{2}s_{2}d_{1}, \gamma(+1) |O|s_{1}u_{1}s_{2}>=A_{1}$
can be described by three diagrams Fig.2(1) with the
quark $ s_{2}$ as a spectator.
Finally, the 6th matrix element on the RHS of the Eq.(A.4) 
$ <d_{2}s_{2}u_{1}, \gamma(+1) |O|s_{1}u_{1}s_{2}>=A_{2}$
can be described by three diagrams Fig.2(2) with the
quark $ s_{2}$ as a spectator. So
$$ <\Sigma^{0}_{\downarrow}, \gamma(+1) |O| \Xi^{0}_{\uparrow} >=
\frac{2}{3 \sqrt{2}}(A_{1}+A_{2}-2A_{3}+4A_{4}) .$$  
\end{enumerate}

\pagebreak
\begin{center}
{\bf Table 1.}  Hyperon weak radiative transitions, experiment~\cite{PDG} 
\vspace{5mm}

\begin{tabular}{|c|c|c|} \hline
Decay                &  BR $(\times 10^{3}) $      
&   Asymmetry           \\ \hline
$\Sigma^{+} \rightarrow p \gamma$        & $1.23 \pm 0.06$    
&  $-0.76 \pm 0.08$     \\ \hline 
$\Sigma^{0} \rightarrow n \gamma$        & $-$                
&       $-$             \\ \hline
$\Lambda^{0} \rightarrow n \gamma$       & $1.63 \pm 0.14$    
&       $-$             \\ \hline
$\Xi^{0} \rightarrow \Lambda \gamma$     & $1.06 \pm 0.16$    
&  $+0.44 \pm 0.44 $    \\ \hline
$\Xi^{0} \rightarrow \Sigma^{0} \gamma$  & $3.56 \pm 0.43$    
&  $+0.20 \pm 0.32 $    \\ \hline
$\Xi^{-} \rightarrow \Sigma^{-} \gamma$  & $0.128 \pm 0.023$  
&  $+1.0 \pm 1.3$       \\ \hline
\end{tabular}
\end{center}

\vspace{10mm}

\begin{center}
{\bf Table 2.}  Hyperon weak radiative PV transitions, theory,
$SU(3)_{f}$ model and single-quark diagram contributions 
\vspace{5mm}

\begin{tabular}{|c|c|c|c|c|} \hline
Decay  &  in \cite{Sharma} & in \cite{Hara} 
& from Eqs.(\ref{new})  & from Eq.(\ref{wnc})\\ \hline
$ \Sigma^{+}\rightarrow p \gamma$  & $-b/3 $ & $0$ 
& $c^{T}$ & $-F+D$ \\ \hline 
$ \Sigma^{0}\rightarrow n \gamma$  & $b/3\sqrt{2}  $
& $a^{d}/\sqrt{2} $
& $a^{T}/\sqrt{2} $ & $(F-D)/\sqrt{2}$  \\ \hline
$\Lambda^{0}\rightarrow n \gamma$  & $3b/\sqrt{6}$ 
& $a^{d}/\sqrt{6}  $
& $a^{T}/\sqrt{6} $ & $-(3F+D)/\sqrt{6}$ \\ \hline
$\Xi^{0}\rightarrow \Lambda \gamma$  & $b/\sqrt{6}$ 
& $-a^{d}/\sqrt{6}$
& $a^{T}/\sqrt{6} $ & $(3F-D)/\sqrt{6}$  \\ \hline
$\Xi^{0}\rightarrow \Sigma^{0} \gamma$  & $ -5b/3\sqrt{2}$ 
& $ -a^{d}/\sqrt{2} $
& $ a^{T}/\sqrt{2}$ & $ -(F+D)/\sqrt{2}$ \\ \hline
$ \Xi^{-}\rightarrow \Sigma^{-} \gamma$  & $5b/3$ & $0$ 
& $ b^{T}$ & $F+D$\\ \hline
\end{tabular}
\end{center}
\vspace{10mm}

\begin{center}
{\bf Table 3.}  Hyperon weak PV radiative transitions, theory, 
2-quark diagram contributions 
\vspace{5mm}

\begin{tabular}{|c|c|c|c|} \hline
Decay &  in \cite{Zen} & in \cite{Sharma} 
& from Eqs.(\ref{AAA})  \\ \hline
$ \Sigma^{+}\rightarrow p \gamma$  & $-\frac{5+\epsilon}{9\sqrt{2}}b$ 
& $ \frac{2}{9}[-3-2X+\zeta(3+X)]$ 
& $ \frac{2}{3}(-2A_{1}+A_{2}-2A_{3}+A_{4})$  \\ \hline 
$ \Sigma^{0}\rightarrow n \gamma$  & $-\frac{1-\epsilon}{18}b  $
& $ \frac{2}{9\sqrt{2}}[-2X+\zeta(-3+X)] $
& $ \frac{2}{3\sqrt{2}}(A_{1}-2A_{2}-2A_{3}+A_{4}) $   \\ \hline
$ \Lambda^{0}\rightarrow n \gamma$  & $\frac{3+\epsilon}{6\sqrt{3}}b$ 
& $ \frac{2}{3\sqrt{6}}[-2+\zeta(-3+X)] $
& $ \frac{2}{\sqrt{6}}(A_{1}-2A_{2}-A_{4}) $ \\ \hline
$\Xi^{0}\rightarrow \Lambda \gamma$  & $-\frac{2+\epsilon}{9\sqrt{3}}b$ 
& $ \frac{2}{3\sqrt{6}}[1-2\zeta] $
& $ \frac{2}{\sqrt{6}}(A_{1}-A_{2})$  \\ \hline
$ \Xi^{0}\rightarrow \Sigma^{0} \gamma $  & $ \frac{1}{3}b $ 
& $ \frac{2}{9 \sqrt{2}}[-3-2X-2\zeta X] $
& $ \frac{2}{3 \sqrt{2}}(A_{1}+A_{2}-2A_{3}+4A_{4})$ \\ \hline
$ \Xi^{-}\rightarrow \Sigma^{-} \gamma$  & $0$ & $0$ & $0$ \\ \hline
\end{tabular}
\end{center}

\newpage
\thispagestyle{empty}
\begin{figure}
\epsfig{file=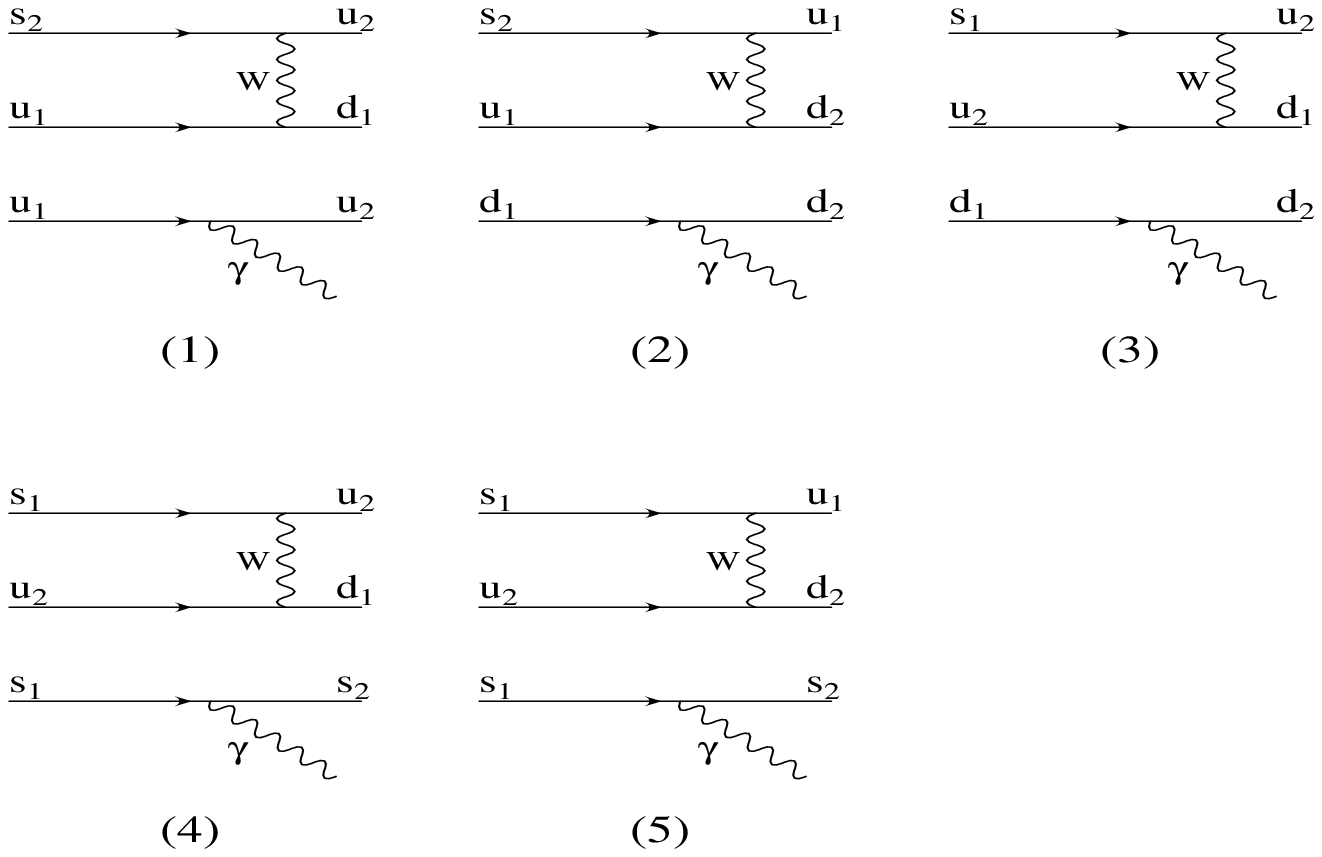}

\vspace{1cm}

\caption{Three-quark diagrams without spectator quark 
($q_{1}$ means $q_{\uparrow}$, $q_{2}$ means $q_{\downarrow}$, $q=u,d,s$)}
\end{figure}

\newpage
\thispagestyle{empty}
\begin{figure}
\epsfig{file=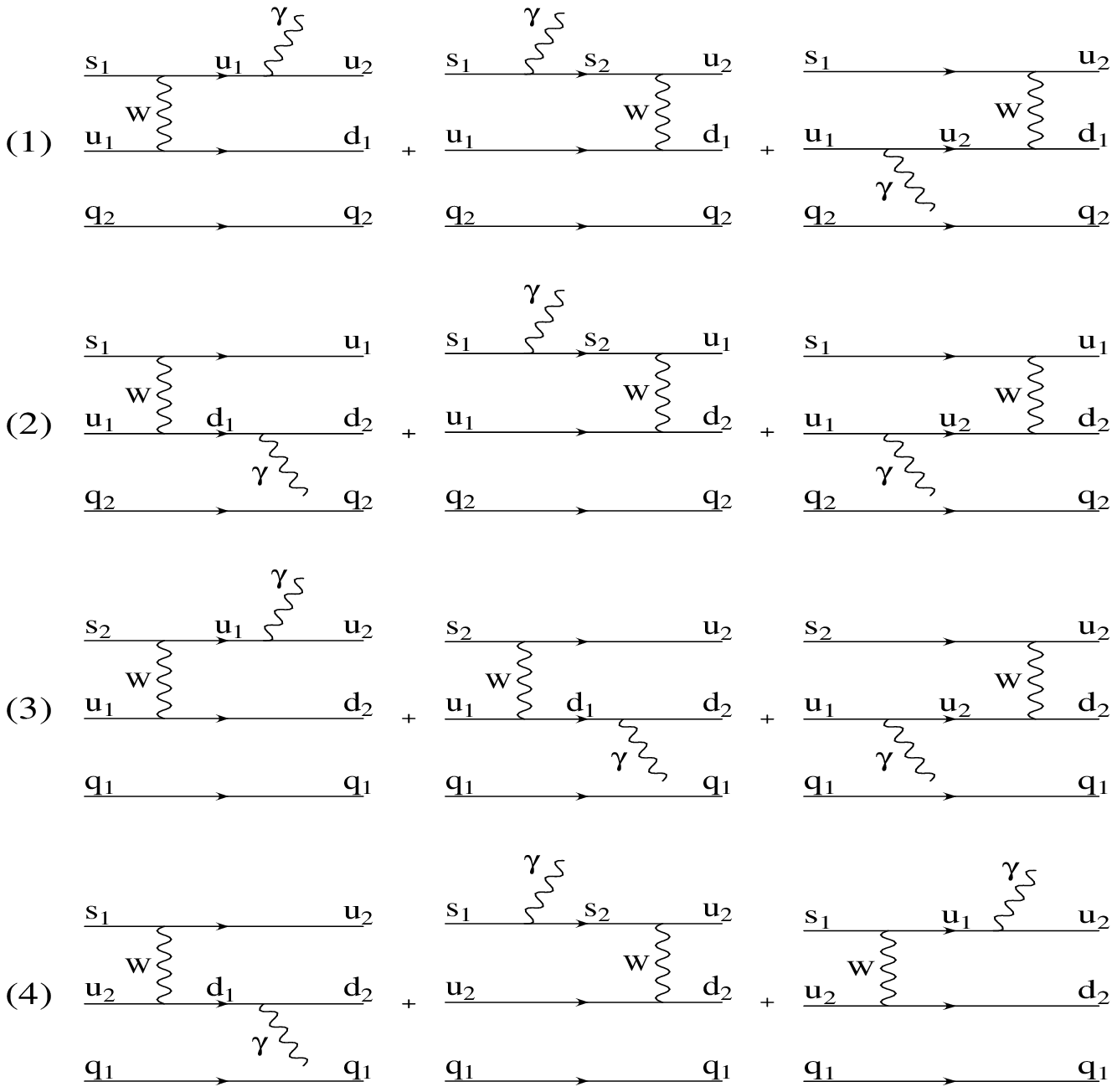}

\vspace{1cm}

\caption{Three-quark diagrams with the third quark $q_{1,2}$ as a spectator 
($q_{1}$ means $q_{\uparrow}$, $q_{2}$ means $q_{\downarrow}$, $q=u,d,s$)}
\end{figure}


\begin{thebibliography}{99}
\bibitem{Behr} R.E.~Behrends, Phys.Rev. {\bf 111}, 1691 (1958)
\bibitem{Furl} G.~Calucci, G.~Furlan, Il Nuovo Cimento {\bf 21}, 679 (1961)
\bibitem{Pati} J.C.~Pati, Phys.Rev. { \bf 130}, 2097 (1963)
\bibitem{Salam} G.~Feldman et al., Phys.Rev. 
               {\bf 121}, 302 (1961)
\bibitem{Nish} M.~Kawaguchi and N.~Nishijima, Prog.Theor.Phys. 
{ \bf 15}, 182 (1956);
C.~Iso and M.~Kawaguchi, Prog.Theor.Phys. { \bf 16}, 177 (1956)
\bibitem{Bo} G.~Quareni et al., Il Nuovo Cimento { \bf 14}, 1179 (1958)
\bibitem{Te} J.~Schneps and Y.W.~Kang, Il Nuovo Cimento { \bf 19}, 1218 (1961)
\bibitem{Pak} R.H.~Graham and S.~Pakvasa, Phys.Rev. {\bf 140}, B1144 (1965)
\bibitem{Hara} Y.~Hara, Phys.Rev.Lett. { \bf 12}, 378 (1964)
\bibitem{Ger} L.K.~Gershwin et al., Phys.Rev. { \bf 188}, 2077 (1969)
\bibitem{PDG} Particle Data Group, The Eur.Phys. J.C { \bf 3}, 613 (1998);
Phys.Rev.{\bf D54}, 1-I (1996)
\bibitem{ZenM} J.~Lach and P.~Zenczykowski, Int.J.Mod.Phys.A {\bf 10},
              3817 (1995)
\bibitem{AG} M.~Ademollo and R.~Gatto, Phys.Rev.Lett. {\bf 13}, 264 (1964)
\bibitem{Ryaz} A.N.~Kamal and Riazuddin, Phys.Rev. {\bf D28}, 2317 (1983)
\bibitem{Sharma} R.C.~Verma and A.~Sharma, Phys.Rev. {\bf D38}, 1443 (1988)
\bibitem{Verma} A.N.~Kamal and R.C.~Verma, Phys.Rev. {\bf D26}, 190 (1982)
\bibitem{Zen} P.~Zenczykowski, Phys.Rev. {\bf D44}, 1485 (1991);
ibid.{\bf D40}, 2290 (1989)
\bibitem{Dmit} V.~Dmitra$\breve{\mbox{s}}$inovi$\acute{\mbox{c}}$, Phys.Rev.{\bf D54}, 5899 (1996)
\bibitem{Gad} N.G.~Deshpande and G. Eilam, Phys.Rev. {\bf D26}, 2463 (1982)
\bibitem{Vasa} N.~Vasanti, Phys.Rev. {\bf D13}, 1889 (1976)
\bibitem{Chesh} V.M.~Dubovik and A.A.~Cheshkov,
                Sov.J.Part.Nucl.{\bf 5}, 318 (1974)
\bibitem{DT} V.M.~Dubovik and V.V.~Tugushev, 
             Phys.Rep.{\bf 187}, 145 (1990)
\bibitem{Bukina} E.N.~Bukina, V.M.~Dubovik, V.S.~Zamiralov,
Phys.Lett. {\bf B449}, 93 (1999)
\bibitem{BDK}   E.N.~Bukina, V.M.~Dubovik, V.E.~Kuznetzov,
            preprints JINR, P2-97-411, P2-97-412, Dubna, Russia, 1997
\bibitem{Bart} G.~Barton, {\it Introduction to Dispersion Techniques in
               Field Theory}, (Benjamin 1965)
\bibitem{Sachs} R.G.~Sachs, {\it Nuclear Theory}, (Cambridge 1953) and 
                 Phys.Rev.Lett.{\bf 13}, 286 (1964)
\bibitem{Morp} G.~Morpurgo, Physics { \bf 2}, 95 (1965); W.Thirring,
Acta Phys.Austr.Suppl.{ \bf 2}, 205 (1965)
\bibitem{Gas} S.~Gasiorowicz, {\it Elementary Particle Physics},
(John Wiley 1966) 
\bibitem{Penguin} S.G.~Kamath,Nucl.Phys.{\bf B198},61(1982);J.O.~Eeg,
Z.Phys.C{\bf 21},253 (1984)
\bibitem{Marshak} R.E.~Marshak, Riazuddin and C.P.~Ryan,
{\it Theory of Weak Interactions in Particle Physics},
(Wiley-Interscience 1965)
\bibitem{Valya} V.M.~Dubovik,V.E.~Kuznetsov, Int.J.Mod.Phys.
{\bf A13}, 5257 (1998)
\bibitem{Fabbri} S.~Bertolini,M.~Fabbrichese and E.~Gabrielli,
Phys.Lett.{\bf B327}, 1361 (1994)
\bibitem{Mendel} G.~Eilam,A.~Ioannissian,R.R.~Mendel and P.~Singer,
Phys.Rev.{\bf D53}, 3629 (1996)
\bibitem{Donoghue} J.F.~Donoghue, Phys.Rev.{\bf D15}, 184 (1976).
\bibitem{DZZ} V.M.~Dubovik, V.S.~Zamiralov and S.V.~Zenkin,
Nucl.Phys.{\bf B182}, 52 (1981)
\bibitem{China} H.-Y.~Cheng, C.-Y.~Cheng, G.-L.~Lin, Y.C.~Lin, T.-M.~Yan, 
H.-L.Yu, Phys.Rev.{\bf D51}, 1199 (1995).
\end{thebibliography}
\end{document}